\documentclass[aps,prc,twocolumn,superscriptaddress]{revtex4-1}

\usepackage{graphicx}
\usepackage{verbatim}
\usepackage{textgreek}

\begin{document}


\title{Observation of T=3/2 Isobaric Analog States in $^9$Be using p+$^8$Li resonance scattering}


\author{C. Hunt}
\affiliation{Department of Physics and Astronomy, Texas A \& M University, 77843 TX}
\affiliation{Cyclotron Institute, Texas A \& M University, 77843 TX}
\author{G.V. Rogachev}

\email[]{rogachev@tamu.edu}
\affiliation{Department of Physics and Astronomy, Texas A \& M University, 77843 TX}
\affiliation{Cyclotron Institute, Texas A \& M University, 77843 TX}
\affiliation{Nuclear Solutions Institute, Texas A \& M University, 77843 TX}

\author{S. Almaraz-Calderon}
\affiliation{Department of Physics, Florida State University, 32306 FL}

\author{A. Aprahamian}
\affiliation{Department of Physics, University of Notre Dame, 46556 IN}

\author{M. Avila}
\affiliation{Argonne National Laboratory, Lemont, 60439 IL}

\author{L.T. Baby}
\affiliation{Department of Physics, Florida State University, 32306 FL}

\author{B. Bucher}
\affiliation{Idaho National Laboratory, Idaho Falls, 83415 ID}
 
\author{V.Z. Goldberg}
\affiliation{Cyclotron Institute, Texas A \& M University, 77843 TX}

\author{E.D. Johnson}
\affiliation{Department of Physics, Florida State University, 32306 FL}

\author{K.W. Kemper}
\affiliation{Department of Physics, Florida State University, 32306 FL}

\author{A.N. Kuchera}
\affiliation{Department of Physics, Davidson College, Davidson, 28035 NC}

\author{W.P. Tan}
\affiliation{Department of Physics, University of Notre Dame, 46556 IN}

\author{I. Wiedenh\"over}
\affiliation{Department of Physics, Florida State University, 32306 FL}


\date{\today}

\begin{abstract}
\begin{description}
\item[Background] Resonance scattering has been extensively used to study the structure of exotic, neutron-deficient nuclei. Extension of the resonance scattering technique to neutron-rich nuclei was suggested more than 20 years ago. This development is based on the isospin conservation law. In spite of broad field of the application, it has never gained a wide-spread acceptance.  
\item[Purpose] To benchmark the experimental approach to study the structure of exotic neutron-rich nuclei through resonance scattering on a proton target.
\item[Method] The excitation function for p+$^8$Li resonance scattering is measured using a thick target by recording coincidence between light and heavy recoils, populating T=3/2 isobaric analog states (IAS) in $^9$Be.
\item[Results] A good fit of the $^8$Li(p,p)$^8$Li resonance elastic scattering excitation function was obtained using previously tentatively known 5/2$^-$ T=3/2 state at 18.65 MeV in $^9$Be and a new broad T=3/2 s-wave state - the 5/2$^+$ at 18.5 MeV. These results fit the expected iso-mirror properties for the T=3/2 A=9 iso-quartet.
\item[Conclusions] Our analysis confirmed isospin as a good quantum number for the investigated highly excited T=3/2 states and demonstrated that studying the structure of neutron-rich exotic nuclei through IAS is a promising approach.
\end{description}
\end{abstract}


\maketitle

\section{\label{Introduction}Introduction}

The understanding of nuclear structure evolution with increasing values of isospin has been the mainstream in contemporary nuclear science for many decades. Development of rare isotope beams provided a major experimental advantage in these studies because simple and well understood reactions, such as nucleon-transfer or Coulomb excitation reactions, could now be used to populate states in exotic nuclei over a range of isospins far removed from the valley of stability.

Resonance scattering with rare isotope beams using the thick target inverse kinematics (TTIK) approach \cite{artemov90} is a particularly powerful technique that has been extensively used to establish the level structure of exotic proton-rich nuclei. Many nuclei have been studied this way over the last 25 years, including the first observations of ground states in several unbound nuclei ($^{10}$N \cite{hooker17}, $^{11}$N \cite{axelsson96}, $^{14}$F \cite{goldberg10}, $^{15}$F \cite{peters03}, $^{18}$Na \cite{assie11}). Advantages of this technique, such as high efficiency, excellent energy resolution ($\sim$20 keV in c.m.), and a well understood reaction mechanism described by R-matrix theory \cite{lane58} made it a technique of choice when applicable. However, application of the resonance scattering approach has been limited primarily to the proton-rich side of the nuclear chart. 

\begin{figure*}[t]
   \includegraphics[width=\textwidth]{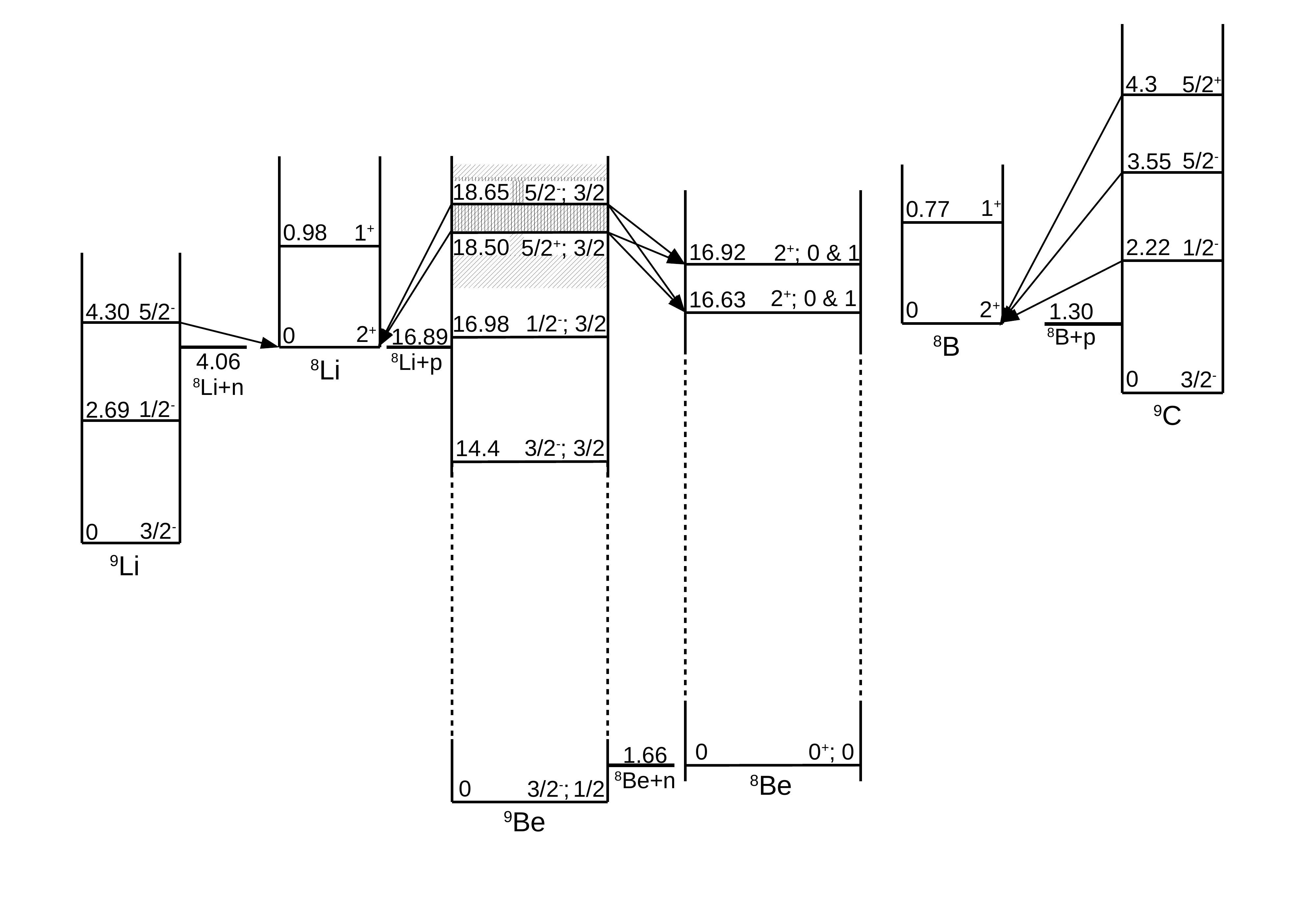}
     \caption{The level structure of the A=9, T=3/2 iso-quartet with levels for $^8$Li, $^9$Li, $^8$Be, $^9$Be, and $^8$B from \cite{tilley04} and $^9$C from  \cite{rogachev07,brown17,hooker19}.}
     \label{fig:levels}
\end{figure*}

Direct extension of TTIK to neutron-rich nuclei, which would involve resonance scattering of rare isotope beams off of neutron targets, is not possible due to lack of the latter. However, one can employ isospin symmetry to study neutron-rich nuclei through the isobaric analog states (IAS) which can be efficiently populated in resonance scattering of neutron-rich ions off of a proton target.  
This approach was first mentioned in an ENAM 1998 conference proceedings \cite{goldberg98}, and originally implemented in $^8$He+p resonance scattering measurements \cite{rogachev03}. The main idea is that while the T=5/2 (T-high) and T=3/2 (T-low) states in $^9$Li are populated in the  $^8$He(T=2)+p(T=1/2) resonance scattering, the T=5/2 (T-high) states would dominate the p+$^8$He excitation function for resonance elastic scattering. This is because only a few isospin allowed decay channels are open for these states, with proton decay back to $^8$He (elastic scattering) and isospin allowed neutron decay to the $^8$Li(T=2,0$^+$), the IAS of $^8$He(g.s.), as the two main decay channels. The most graphic confirmation of this idea was demonstrated in the experiment in which the excitation function for the $^6$He(p,n)$^6$Li(T=1,0$^+$) reaction was measured \cite{rogachev04}. The T=3/2 (T-high) states completely dominate the spectrum of $^7$Li  measured in \cite{rogachev04} while no evidence for T=1/2 (T-low) states was observed. Yet, acceptance of this approach was slow in the community. This is primarily due to concerns associated with the role of the T-low states and the validity of the isospin-symmetry hypotheses for very exotic nuclei deep into the continuum. In addition to the already mentioned experiments, there were two more applications of this approach to study the structure of light, neutron rich nuclei, $^9$He \cite{uberseder16} and $^{13}$B \cite{skorodumov07} and several recent studies in medium mass region \cite{bradt18, imai10, imai12, imai14} that applied the TTIK technique with rare isotope beams to study $^{47}$Ar, $^{68}$Zn, $^{35}$Si, and $^{31}$Mg.

The main goal of this work is to study a benchmark case that can be used to explore the applicability and limitations of the proposed experimental concept for spectroscopic studies of neutron-rich nuclei. A convenient case is the A=9 T=3/2 iso-quartet shown in Fig. \ref{fig:levels}, that consists of \textsuperscript{9}Li, $^9$Be(T=3/2), $^9$B(T=3/2), and $^9$C. Discussion if isospin is a good symmetry for the A=9 iso-quartet dates far back to the time when mass measurements for $^9$C first became available \cite{Trentelman1970}. The structure of \textsuperscript{9}C has been studied recently using resonance scattering and the invariant mass technique \cite{rogachev07,brown17,hooker19} and its low-lying levels are well established now. The lowest states in \textsuperscript{9}Li have been studied with the $^8$Li(d,p) reaction \cite{wuosmaa05} and also with $^7$Li(t,p) \cite{middleton64}. Experimental information on the three lowest T=3/2 states in \textsuperscript{9}Be is also available \cite{tilley04}. Therefore, one can expect that if the T=3/2 states dominate the $^8$Li+p resonance elastic scattering and if isospin is a good symmetry then the excitation function for this reaction can be reasonably well constrained from the already available data. A surprising claim to the contrary was made recently in \cite{leistenschneider18} where analysis of low energy resonances populated in $^8$Li+p scattering revealed significant isospin mixing for this specific case. We have performed kinematically complete measurements of the excitation functions for $^8$Li+p elastic and inelastic scattering in the c.m. energy range from 1.46 MeV to 2.3 MeV which corresponds to $^9$Be excitation energy range from 18.35 MeV to 19.19 MeV. Combining spectroscopic information already available for the A=9 T=3/2 iso-quartet, two T=3/2 states are expected at these energies in the spectrum of $^9$Be. It is the 5/2$^-$ state at 18.65 MeV and the 5/2$^+$ state at around 18.5 MeV (see detailed discussion in sec. \ref{Analysis}). The R-matrix analysis of the $^8$Li+p excitation functions measured in this work conclusively demonstrates that these two T=3/2 states provide a perfect description of the experimental data, lending strong support to the experimental approach proposed 22 years ago \cite{goldberg98}. No evidence for isospin mixing in $^9$Be has been observed.

\begin{figure}[t]
   \includegraphics[width=\columnwidth]{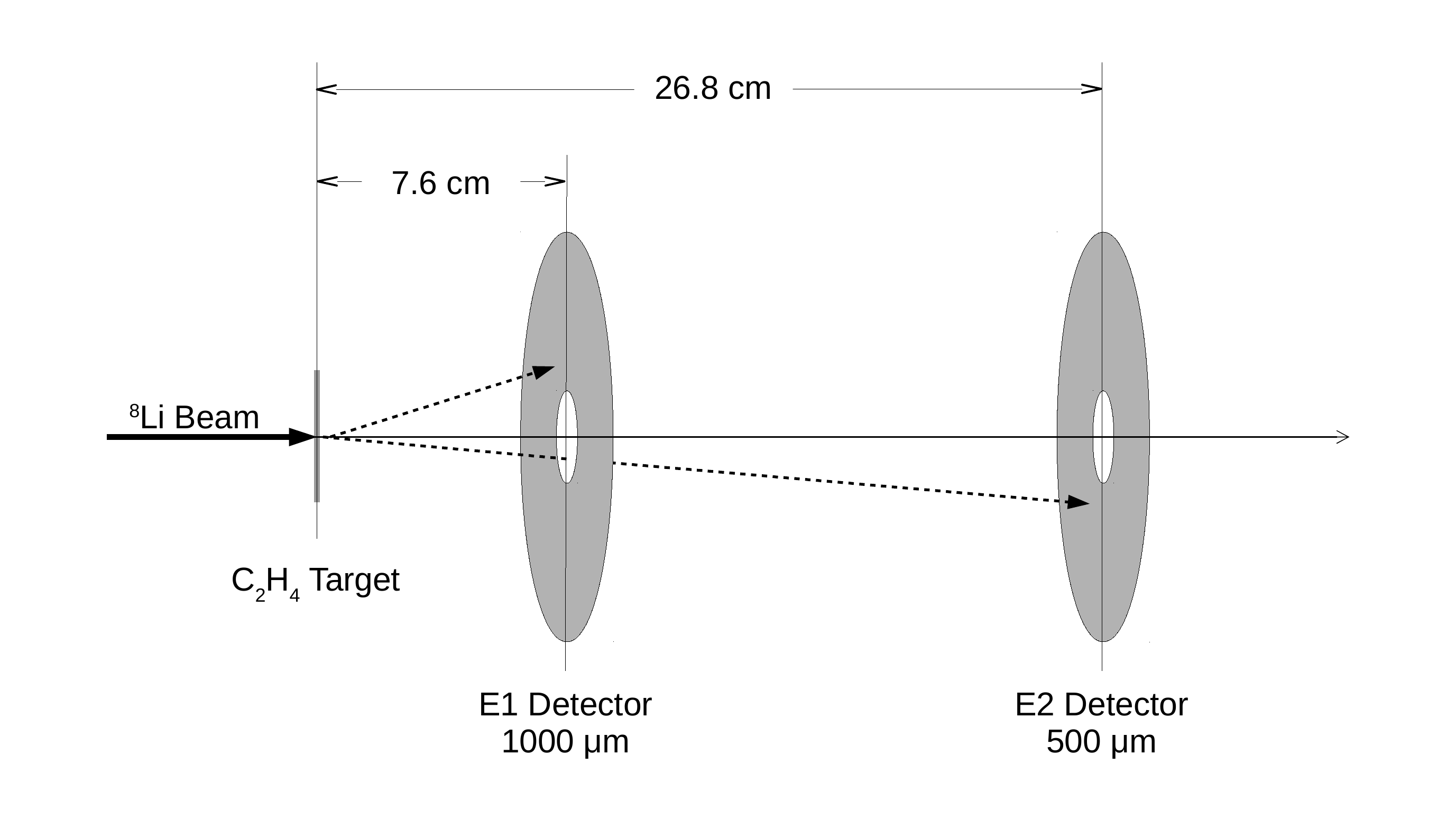}
     \caption{Schematic view of the experimental setup. Two Micronsemiconductors Ltd. annular double-sided strip detectors of ``S2'' type were installed after the polyethylene target. Detectors were centered on the beam axis as shown. Position sensitive E1 and E2 detectors were used to measure total energy and hit location of the light and heavy recoils from the $^8$Li+p reactions, respectively.}
     \label{fig:8liExpSetup}
\end{figure}

\section{\label{Experiment}Experiment}

This experiment was carried out at the RESOLUT \cite{RESOLUT} radioactive nuclear beam facility at the John D. Fox Superconducting Accelerator Laboratory at Florida State University using the hybrid Thick/Thin Target in Inverse Kinematics approach \cite{rogachevttik,Almaraz2011}. In this approach, the target is thick enough for the beam particles to lose a significant fraction of their energy, but thin enough for the heavy recoil particles to exit the target and be detected. A radioactive $^{8}$Li beam (t$_{1/2}$ = 838 ms) was produced using the $^{2}$H($^{7}$Li,$^{8}$Li)$^{1}$H reaction. The primary $^{7}$Li beam was accelerated by a 9 MV FN tandem Van de Graaff accelerator followed by a linear accelerator booster to kinetic energies of 27 MeV and 23.5 MeV (two beam energies were used in this experiment). The primary target was a liquid-nitrogen-cooled, 4 cm long deuterium gas cell with pressure of 400 Torr and 2.5 $\mu$m thick Havar entrance and exit windows.  The secondary $^{8}$Li beam was momentum selected, bunched and separated from other contaminants by the superconducting resonator, and the quadrupole and dipole magnets of the RESOLUT separator. The composition of the radioactive beam was 95 $\%$ $^{8}$Li and 5 $\%$ $^{7}$Li contaminant at the secondary target position. The typical intensity of the $^{8}$Li beam was $\approx$ 2$\times$10$^{4}$ pps. We measured the excitation function for $^{8}$Li+p in the energy region between 1.46 and 2.3 MeV in the c.m. system. The proton decay threshold in $^9$Be is at excitation energy of 16.888 MeV (Fig. \ref{fig:levels}), so we covered the excitation energy range from 18.35 to 19.19 MeV. The light and heavy reaction residues were measured in coincidence. Two $^{8}$Li-beam energies were used in this experiment: 22.0 MeV and 18.6 MeV. A polyethylene (C$_{2}$H$_{4}$) target thickness was optimized for each beam energy to ensure that both light and heavy recoils get out of the target with enough energy to be detected. By carefully choosing the combination of the beam energy and the target thickness, it was possible to measure the continuous excitation functions for $^{8}$Li+p elastic scattering from 1.46 to 2.3 MeV in the c.m. system in just two beam energy steps. The thickness of the polyethylene target was 4.13 mg/cm$^{2}$ for the $^{8}$Li beam energy of 22.0 MeV. Two different target thicknesses, 4.13 mg/cm$^{2}$ and 2.75 mg/cm$^{2}$, were used with the 18.6 MeV $^{8}$Li beam energy.

Two Micron Semiconductors Ltd. \cite{micron} annular silicon strip detectors of the S2 type were installed downstream of the target along the beam axis. A schematic view of the experimental setup is shown in Fig. \ref{fig:8liExpSetup}. The S2 detectors have annular geometry and they consist of 48 rings on one side, that were combined into groups of three for a total of sixteen channels, and sixteen segments on the other side. The first S2 detector (E1), which had a thickness of 1000 $\mu$m, was placed at 7.6 cm to measure light recoils, covering an angular range from 8.2$^{\circ}$ to 24.7$^{\circ}$ in the laboratory reference frame. The second S2 detector (E2), which had a thickness of 500 $\mu$m for measuring heavy recoils, was located at 26.8 cm downstream from the target, covering an angular range from 2.4$^{\circ}$ to 7.4$^{\circ}$.

\section{\label{Results}Experimental Results}

In this experiment we measured the complete kinematics for the binary reactions. The trigger was set to coincidence mode between the E1 and E2 detectors. Only those events that produced signals in both near and far S2 detectors simultaneously (within 100 ns) were recorded. In addition to measuring energies of heavy and light recoils, direction of the momentum vectors can be recovered for both particles from the location of the hits, extracted from the double-sided annular strip detectors. Coincidence between light and heavy recoils in two S2s and complete kinematics allows for unambiguous and background-free identification of the binary reaction channels.

\begin{figure}[t]
	\includegraphics[width=\columnwidth]{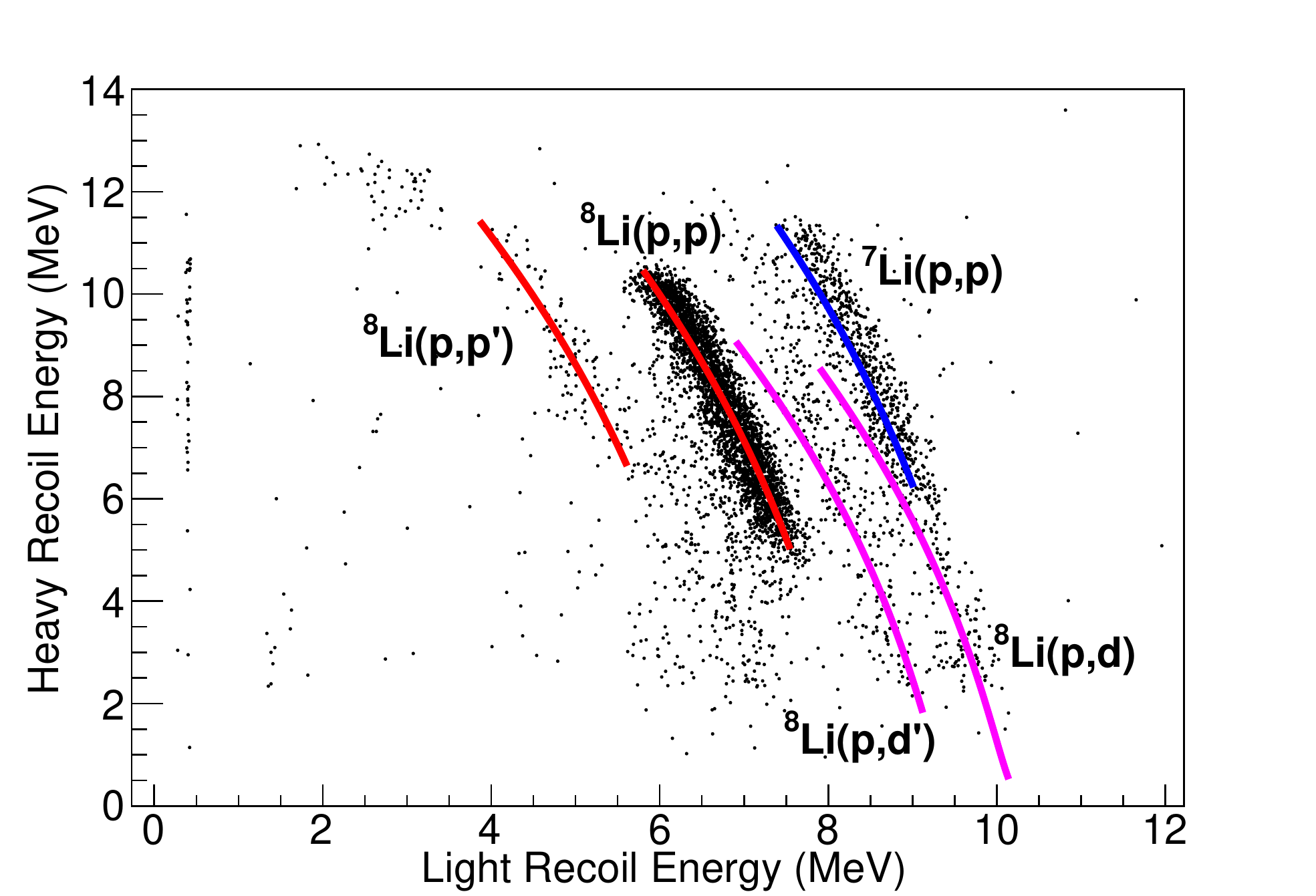}
	\caption{\label{fig:E1E2Plot} Scatter plot of energy measured in the heavy recoil detector (E2) plotted against the energy measured in the light recoil detector (E1). The curves show calculated heavy recoil energy vs light recoil energy correlation for binary reactions of $^8$Li beam on the proton target. The red curves represent $^8$Li+p elastic and inelastic scattering populating the first excited state in $^8$Li. The blue curve is the $^7$Li+p elastic scattering and the magenta curves are the $^8$Li(p,d)$^7$Li(g.s.) and $^8$Li(p,d')$^7$Li(0.48 MeV) reactions.}
\end{figure}

Fig. \ref{fig:E1E2Plot} shows a 2D identification scatter plot. Energy deposited by the heavy ion in the E2 detector is plotted versus energy deposited by the light ion in the E1 detector. The calculated kinematics curve for various reaction channels is shown for comparison. The most intense group is due to the $^8$Li+p elastic scattering. This is not surprising, of course, because the cross section for elastic scattering is high and the geometry of the experimental setup was optimized for this channel. Since there was a 5\% contamination of $^7$Li in the secondary beam we also expect to see $^7$Li(p,p) elastic scattering, which is clearly visible in Fig. \ref{fig:E1E2Plot} at higher total energy. This is just as expected because the kinetic energy of the $^7$Li beam was higher than that of the $^8$Li it produced. There are three more reaction channels that can be identified in Fig. \ref{fig:E1E2Plot}: inelastic scattering, $^8$Li(p,p'), populating the first excited state in $^8$Li (the 1$^+$ at 0.98 MeV), and the $^8$Li(p,d) reactions populating the ground and the first excited states in $^7$Li. Statistics are very low for the $^8$Li(p,p') inelastic scattering, but it still carries useful information. It indicates that the cross section for inelastic scattering is smaller than the cross section for elastic scattering by a factor of 30 and therefore this channel can be neglected in the R-matrix analysis described in sec. \ref{Analysis}. The $^8$Li(p,d)$^7$Li(g.s.) reaction channel was used to verify the overall normalization, which was obtained using the ratio of the $^8$Li ions to the primary beam current. We verified that the $^8$Li(p,d)$^7$Li(g.s.) reaction cross section measured in this experiment is in good agreement with the cross section for the time-reverse $^7$Li(d,p)$^8$Li reaction measured in \cite{lombaard74} and converted using the detailed balance principle.

\begin{figure}[t]
	\includegraphics[width=\columnwidth]{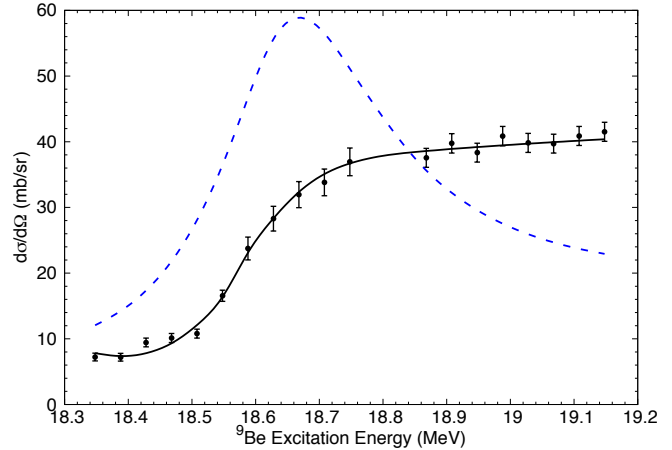}
	\caption{\label{fig:CSFits} Excitation function for $^8$Li+p elastic scattering for an angular range between 138$^{\circ}$ and 155$^{\circ}$ in c.m. The solid curve is the best $R$-matrix fit with T=3/2 5/2$^-$ at 18.65 MeV and T=3/2 5/2$^+$ at 18.5 MeV states in $^9$Be with parameters shown in Table \ref{tab:rmatrix}. The blue dashed curve is the $R$-matrix calculation with the T=3/2 5/2$^-$ state at 18.65 MeV only.}
\end{figure}

Gating on the $^8$Li(p,p) elastic scattering using the 2D scatter plot shown in Fig. \ref{fig:E1E2Plot} and calculating the c.m. energies at the interaction point for each event using energies and scattering angles of both light and heavy recoils (see \cite{rogachevttik}) the excitation function for $^8$Li+p resonance elastic scattering was obtained (Fig. \ref{fig:CSFits}). This excitation function includes c.m. angles from 138$^{\circ}$ to 155$^{\circ}$ in c.m. The smallest and largest angles were excluded to avoid geometric effects of loosing coincidence between the light and heavy reaction residues due to angular divergence and finite spot size of the beam. Energy resolution is dominated by the intrinsic energy resolution of the E1 detector and is about 30 keV in c.m.

\begin{table*}[t]
\caption{\label{tab:rmatrix} Best fit $R$-matrix parameters for the T=3/2 states in $^9$Be with channel radius of 4.5 fm and $\gamma^2_{sp}$=1.25 MeV. E$_{ex}$ is an excitation energy in $^9$Be, E$_{\lambda}$ is an energy eigenvalue, $\Gamma$ is a total width and S is a spectroscopic factor. Natural boundary condition is used so that it is equal to the shift function calculated at the resonance energy, making E$_{\lambda}$ equal to p+$^8$Li c.m. energy. The parameters that were varied in the $R$-matrix fit are boldfaced. The remaining values were recalculated based on the values of the boldfaced parameters and Eq. (1)-(4). The spectroscopic factor for the 5/2$^+$ state was set to unity.}
\begin{ruledtabular}
\begin{tabular}{lllllllll}
J$^\pi$  & E$_{ex}$ & E$_{\lambda}$ & $\Gamma$ & S & $\gamma^2_p$ & $\gamma^2_{n(16.626)}$ & $\gamma^2_{n(16.922)}$ \\
   & MeV & MeV & keV &  & keV & keV & keV \\    
\hline 
${\frac{5}{2}}^-$  & 18.65(2) & {\bf 1.76(2)}  & 350(40)  & 1.2(1) & {\bf 510(50)}  & 410 & 610 \\
${\frac{5}{2}}^+$  & 18.5(1) & {\bf 1.6(1)}  & 1500 & 1.0 & 410  & 330 & 490\\
\end{tabular}
\end{ruledtabular}
\end{table*}

\section{\label{Analysis}R-matrix Analysis}

Analysis of the excitation function for $^8$Li+p elastic scattering was performed with the $R$-matrix code MinRMatrix \cite{johnson08}. As was mentioned in the introduction section, some spectroscopy information on the level structure of $^9$Li, $^9$C, and T=3/2 states in $^9$Be in the relevant energy region is available. Therefore, many $R$-matrix parameters can be fixed {\it a priori} for this system. Two T=3/2 states at 14.3922 and 16.9752 MeV are well known in $^9$Be \cite{tilley04}. These are the IAS of the ground (3/2$^-$) and the first excited (1/2$^-$) states of $^9$Li and $^9$C. Note that these states are very narrow - 380 eV each \cite{tilley04}. This is because the isospin allowed nucleon decay channels are energetically forbidden and the resonance widths are dominated by small isospin violating admixtures. The third T=3/2 state is a tentative 5/2$^-$ at 18.65(5) MeV \cite{tilley04} and it is a rather broad resonance ($\sim$300 keV) because the isospin allowed proton and neutron decays are open for this state (see Fig. \ref{fig:levels}). There is a good reason to assume that the 5/2$^-$ spin-parity assignment is correct. The 5/2$^-$ state in $^9$C has been clearly identified at an excitation energy of 3.6 MeV and well characterized as nearly a single particle state in three recent experiments \cite{rogachev07,brown17,hooker19}. Therefore, one is justified to use a simple potential model to predict the Thomas-Ehrman \cite{TE51} shift between the T=3/2, A=9 isobars for this state. Using conventional parameters for the Woods-Saxon potential with R=1.25$\times^3\sqrt{8}$=2.5 fm and $a=0.65$ fm and adjusting the depth to reproduce the 3.6 MeV excitation energy of the 5/2$^-$ in $^9$C, one gets an excitation energy of 5/2$^-$ in $^9$Li at 4.26 MeV. This is less than 40 keV different from the known tentative 5/2$^-$ state at 4.296 MeV in $^9$Li \cite{tilley04}. Using excitation energies of the 5/2$^-$ in $^9$Li and $^9$C an excitation energy of the T=3/2 5/2$^-$ IAS in $^9$Be can be estimated at 18.5 MeV. Therefore we expect to observe a single-particle T=3/2 5/2$^-$ state in the measured excitation energy region - between 18.35 and 19.19 MeV. Moreover, its $R$-matrix parameters can be tightly constrained by the fact that neutron decay to the T=0 states in $^8$Be should be strongly suppressed due to the isospin conservation. We set the reduced widths associated with these decays to zero. The reduced widths for neutron decay to the isospin mixed T=0+1 states at 16.626 and 16.922 MeV in $^8$Be and proton decay to $^8$Li(g.s.) are defined by the isospin Clebsch-Gordan coefficients, the nearly unity spectroscopic factor of the 5/2$^-$ state \cite{rogachev07,hooker19} and the known isospin mixture of the T=0+1 2$^+$ states in $^8$Be \cite{wiringa13}. They are given by the equations below:

\begin{eqnarray}
{\gamma}^2_p & = & S\gamma^2_{sp} \left ( C{}^{1}_{1}{}^{\,\,\,\,\frac{1}{2}}_{-\frac{1}{2}}{}^{\frac{3}{2}}_{\frac{1}{2}} \right )^2 \\
{\gamma}^2_n & = & S\gamma^2_{sp} \left ( C{}^{1}_{0}{}^{\frac{1}{2}}_{\frac{1}{2}}{}^{\frac{3}{2}}_{\frac{1}{2}} \right ) ^2 \\
{\gamma}^2_{n(16.626)} & = & \gamma^2_n \times 0.4 \\
{\gamma}^2_{n(16.922)} & = & \gamma^2_n \times 0.6,
\end{eqnarray} 

where $\gamma^2_{sp}$ is the single particle reduced width which was set to 1.25 MeV to reproduce the single particle width of a p-wave resonance calculated with the potential model mentioned above at an $R$-matrix channel radius of 4.5 fm. The boundary condition was set equal to the shift function calculated at the resonance energy. Using considerations above, all $R$-matrix parameters for the T=3/2 5/2$^-$ state at 18.65(5) MeV in $^{9}$Be are constrained. 

The $R$-matrix calculations that include only the T=3/2 5/2$^-$ state at 18.65 MeV are shown in Fig. \ref{fig:CSFits} with a dashed blue curve. Parameters for the 5/2$^-$ state are given in Table \ref{tab:rmatrix} and are consistent with  \cite{tilley04}. Obviously, the dashed blue curve does not reproduce the experimental data. Rather, one more T=3/2 state needs to be included. A very broad, purely single-particle $\ell=0$ 5/2$^+$ state has been observed in $^9$C at around 4 MeV excitation energy \cite{hooker19}. Its IAS should be located at around 18.7 MeV in $^9$Be. The single-particle nature of this state in $^9$C allows one to fix the spectroscopic factor to unity and calculate the reduced width using Eq. (1)-(4). To produce the final fit we allowed the excitation energies of the 5/2$^+$ and 5/2$^-$ states to vary. We also allowed variation of the total width of the 5/2$^-$ state but we kept the ratio of the reduced widths fixed, as defined by eq. (1)-(4). The best three-parameter fit is shown in Fig. \ref{fig:CSFits} as a black solid curve and the best fit parameters are given in Table \ref{tab:rmatrix}. The normalized $\chi^2$ of the best fit is 0.98. The best fit parameters for the 5/2$^-$ state are close to the expected values. The excitation energy of 18.5 MeV for the 5/2$^+$ state is in agreement with the predictions of the potential model discussed in \cite{hooker19}, which works well for the broad 2s1/2 $\ell=0$ scattering states in $^8$B, $^9$C, and $^{10}$N and predicts that the 5/2$^+$ partial wave should peak at around 1.8 MeV of p+$^8$Li c.m. energy (18.7 MeV). The uncertainties for the fitted parameters were established using the Monte Carlo technique, which randomly varied all three fitting parameters simultaneously and accepted only those sets that resulted in $\chi^2$ values within 90\% confidence level. 


For completeness we note that while proton decay of the T=3/2 states in $^9$Be to the first excited state in $^8$Li (1$^+$ at 0.98 MeV) is energetically possible, it is strongly suppressed by the penetrability factors. We have observed events associated with the inelastic scattering (see Fig. \ref{fig:E1E2Plot}), but the cross section was a factor of 30 smaller, therefore inelastic scattering cannot have significant influence on the elastic scattering cross section and was excluded from the $R$-matrix fit to reduce the number of free parameters. Also, the 5/2$^-$ state has two sets of reduced widths - one for channel spin 3/2 and one for channel spin 5/2. As it was discussed in \cite{rogachev07,hooker19}, channel spin 5/2 should dominate and we have excluded the reduced widths associated with the channel spin 3/2. An excellent agreement between the three-parameter $R$-matrix fit and the experimental data validates these approximations.	

\section{\label{Conclusion}Conclusion}

The excitation function for $^8$Li+p resonance elastic scattering was measured in the energy range that corresponds to the range between 18.35 MeV and 19.19 MeV excitation energy in $^9$Be. The main goal of these measurements was to provide benchmark data to verify the validity of the isospin symmetry considerations and check if the application of the TTIK approach for spectroscopy studies of neutron rich nuclei with rare isotope beams leads to reliable results. The measured excitation function was perfectly described by the $R$-matrix approach, which included the two T=3/2 states only (5/2$^-$ and 5/2$^+$). Moreover, the best fit reduced widths, total widths, and resonance energies are in agreement with the values expected based on the isospin symmetry considerations and most recent experimental information on the level structure of the T=3/2 A=9 iso-quartet. We confirm that the excited state at 18.65 MeV in $^{9}$Be \cite{tilley04} is indeed a 5/2$^-$ T=3/2 IAS. We have also identified a new broad 5/2$^+$ T=3/2 state at 18.5(1) MeV. It appears that the T=1/2 states play only a minor role in this case. This is probably due to the presence of strong, single-particle T=3/2 resonances which dominate the cross section for $^8$Li+p elastic scattering. It was shown that isospin symmetry considerations are still valid in this case, which features broad states in the continuum. This is encouraging and validates the application of the TTIK method for future spectroscopy studies of neutron-rich nuclei with rare isotope beams.

This work was funded in part by the Department of Energy, Office of Science, under Award No. DE-FG02-93ER40773, and by the National Science Foundation under grant No. PHY-1712953 and No. PHY-1713857.

\bibliography{9BeIAS}

\end{document}